# A Kinetic Model for Photoswitching of magnetism in the High Spin Molecule [Mo(IV)(CN)$_2$(CN-Cu(II)(tren))$_6$](ClO$_4$)$_8$

Rajamani Raghunathan[1,#], S. Ramasesha[1,*], Corine Mathonière[2,$] and Valérie Marvaud[3,&]

[1]Solid State and Structural Chemistry Unit, Indian Institute of Science, Bangalore 560 012, India.

[2]Institut de Chimie de la Matière Condensée de Bordeaux, No. 87, UPR CNRS 9048, Université de Bordeaux 1, Ave Dr. Schweitzer, 33608 Pessac Cedex, France.

[3]Laboratoire de Chimie Inorganique et Matériaux Moléculaires, CNRS UMR-7071, Université Pierre et Marie Curie 75252 Paris Cedex 05, France.

Email: [#] rajamani@sscu.iisc.ernet.in; [*] ramasesh@sscu,iisc,ernet.in; [$] mathon@icmcb-bordeaux.cnrs.fr; [&] marvaud@ccr.jussieu.fr

## ABSTRACT

The heptanuclear complex [Mo(IV)(CN)$_2$(CN-CuL)$_6$]$^{8+}$ exhibits photomagnetism. An earlier microscopic model showed that the transition dipole moments for excitation in different spin manifolds are similar in magnitude. In this paper, we attribute photomagnetism to the long lived S=3 charge transfer excited state for which there appears to be sufficient experimental evidence. We model the photomagnetism by employing a kinetic model which includes internal conversions and intersystem crossings. The key feature of the model is assumption of the existence of two kinds of S=3 states: one which has no direct pathway for internal conversion and the other characterized by slow kinetics for internal conversion to the low-energy states. The trapped S=3 state can decay via a thermally activated barrier to the other S=3 state. The experimental temperature dependence of magnetization plot is fitted using rate constants with Arrhenius dependence. The two different experimental $\chi_M T$ vs. $T$ curves obtained with different irradiation times are fitted with our model. Our studies show that the photomagnetism in these systems is governed by kinetics and not due to differences in oscillator strengths for excitation of the different spin states.

## 1. Introduction

In recent years there has been much excitement in the area of molecular magnetism due to the discovery of photomagnetism in several systems [1]. The early system found to exhibit this phenomenon was alkali metal doped Prussian Blue analogues containing Fe and Co ions [2]. For example, Rb$_{0.54}$Co$_{1.21}$[Fe(CN)$_6$].17H$_2$O system, consisting of about 80% diamagnetic Co(III)-NC-Fe(II) bridges, some magnetic pairs Co(II)-NC-Fe(III) and about 17% Fe vacancy, is mostly nonmagnetic in the dark (nonirradiated) state [3]. Upon irradiation with red laser light, this system becomes ferrimagnetic. The mechanism for the photoinduced magnetic transformation is well understood from studies carried out on systems with different concentrations of alkali metal doping and is attributed to a charge transfer excitation from [Co(III),Fe(II)] to [Co(II),Fe(III)] brought about by the incident laser light [4]. Both Co and Fe ions in the oxidation states Co(III), Fe(II) are diamagnetic while in the Co(II), Fe(III) oxidation states they both are paramagnetic, with spins 3/2 and 1/2 respectively. The magnetic ion pairs thus created experience antiferromagnetic superexchange interaction through the cyanide bridge, resulting in ferrimagnetic ordering at low-temperatures. Although, individual ground states of Co(II) and Fe(III) ions are magnetic, the excited [Co(II),Fe(III)] system which is formed by photoirradiation should be nonmagnetic, as required by the spin selection rules. The nonmagnetic state then undergoes intersystem crossing to a magnetic metastable state. The metastable state is associated with large geometry changes in the unit cell relative to the normal state is evident indirectly from the fact that a unit cell without vacancy, such as $Cs_4Co_4(III)[Fe(II)(CN)_6]$ is not photomagnetic and that Fe vacancy is essential for observing photomagnetism [3].

Long-range magnetic order is also observed upon laser irradiation in the $Cu_2$/Mo system $Cu_2(II)Mo(IV)(CN)_8 \cdot 7H_2O$, which is paramagnetic in the dark state [5]. The mechanism for photomagnetism in this system is the charge transfer from diamagnetic $4d^2$ Mo(IV) to $3d^9$ Cu(II) brought about by light leading to paramagnetic $4d^1$ Mo(V) and a diamagnetic $3d^{10}$ Cu(I) ions. A subsequent superexchange interaction between photogenerated Mo(V) and the remaining Cu(II) neighbours leads to ferromagnetic coupling. The presence of Cu(I) species in the photoexcited state is confirmed from x-ray spectroscopic studies [6]. An interesting molecular photomagnetic system that results from capping the Cu(II) ions in the Cu(II)-Mo(IV) system using the capping ligand tris(2-aminoethyl)amine (tren), is the heptanuclear complex $[Mo(IV)(CN)_2(CN-Cu(tren))_6](ClO_4)_8$, noted $MoCu_6$ (Fig. 1) [7]. This molecule consists of six non-interacting spin-½ Cu(II) ions and exhibits Curie behavior down to very low-temperatures. However, irradiating this system with 406 nm laser light leads to a charge transfer excitation from Mo(IV) to one of the neighbouring Cu(II) to yield the spin-½ Mo(V) ion and one Cu(I) ion. The superexchange interaction between Mo(V) and the remaining five Cu(II) ions in the molecule is ferromagnetic. This superexchange interaction in the charge-transfer excited state leads to a metastable spin S=3 species as evidenced by the $\chi_M T$ vs. $T$ plot of the sample post irradiation. In this communication, we present a kinetic model for the photomagnetic behavior in the MoCu$_6$ heptanuclear complex. In the next sections, we briefly present new experimental data, then review the earlier results on the microscopic model for this system, and present the kinetic model. We end this paper with a section dealing with the discussion on the results of the model.

**2. Experimental details**

Magnetic experiments were carried out with a Quantum Design MPMS-5S magnetometer working in the dc mode. The two photomagnetic experiments described in this paper were performed with a Kr$^+$ laser coupled through an optical fiber directed into the squid cavity. They were performed in the 5-300K range with a magnetic field of 5000 G. For the two experiments, the powder sample of [Mo(IV)(CN)$_2$(CN-Cu(II)(tren))$_6$](ClO$_4$)$_8$ was laid down as a thin layer on the sample holder. The output of the fiber was located at a distance of 5 cm from the sample. Due to the small mass of the sample, the SQUID centering deviates significantly with the temperature. In order to correct the SQUID signal, the data have been corrected by measuring the sample before irradiation and comparing the data with the dc measurements done on 20 mg of the sample. The deduced empirical correction has been applied to all the data collected during a complete experimental run. The compound was irradiated continuously using the multiline 406-415 nm under a magnetic field of 0.5 Tesla at 10K, for one hour for experiment A, and 7 hours for experiment B. The power of the laser light received by the sample was 7 mW/cm$^2$. The post irradiation $\chi_M T$ vs. $T$, ($\chi_M$ is the paramagnetic molar magnetic susceptibility and T the temperature in K) plots have been measured in the dark with a warming rate of 2K/min.

Figure 2 shows the $\chi_M T$ vs. $T$ plots before irradiation, after 1 hour of irradiation and 7 hours of irradiation. Before irradiation, the system behaves as a nearly perfect paramagnet in the 50-300K with a Curie constant of 2.4 cm$^3$K/mol, which is in agreement with the expected value (2.25 cm$^3$K/mol) for the presence of six Cu(II) ($S$ = 1/2, C = 0.375 cm$^3$K/mol with $g_{Cu}$ = 2). Below 50 K, the decrease of $\chi_M T$ when the temperature decreases suggests the presence of very weak antiferromagnetic interactions between spin carriers as expected from the crystal structure that shows that all the spin centers are separated by diamagnetic NC-Mo(IV)-CN bridges [6]. In the case wherein the duration of irradiation was one hour, the $\chi_M T$ at 10K increased from its initial value of 2.18 cm$^3$K/mol to 3.55 cm$^3$K/mol. As shown in Fig. 2, the $\chi_M T$ vs. T is qualitatively modified, with the appearance of a maximum of 3.7 cm$^3$K/mol at 20K. Above 20K, the $\chi_M T$ decreases with increase in temperature, and recovers its initial paramagnetic value near room temperature. In the case where duration of irradiation is 7 hours, the $\chi_M T$ at 10K increased from its initial value of 2.18 cm$^3$K/mol to 3.98 cm$^3$K/mol. The $\chi_M T$ vs. T plot shows the maximum of 4.5 cm$^3$K/mol at 20K. Above 20K, the $\chi_M T$ decreases with increase in temperature, and as in the earlier case recovers its initial value near room temperature. These post-irradiation curves show that the photo-excited state of the system has a higher magnetic moment than the initial state, in agreement with the formation of Mo(V) S=1/2 due to charge transfer excitation and concomitant ferromagnetic interaction with the five remaining Cu(II) centers, $S_{Cu}$=1/2 [6]. Furthermore, for both experiments, we have verified that the system relaxes to its original state, to recover the $\chi_M T$ vs. T plot before irradiation, proving the reversibility of the phenomenon.

## 3. Nature of Excited States in MoCu$_6$

A minimal model for describing the active electrons in the photomagnetism of MoCu$_6$ involves one orbital on each Cu(II) ion which is occupied by the unpaired electron and two orbitals on the Mo(IV) ion; the lower energy orbital which is doubly occupied in the ground state, and the other higher energy orbital which is empty in the ground state. The electron configuration in the ground state of the system is shown pictorially in Fig. 3a. The energy gap $\Delta$ between the two 4d orbitals considered in the model is expected to be small ~ 0.2eV and arises due to deviation of the symmetry from a regular cubic field. In the ground state of the molecule, there is no exchange interaction between the Cu and Mo sites and there are no exchange pathways between any two copper sites, all the $2^6$ spin orientations are degenerate and we obtain the Curie behavior. Indeed, the $\chi_M T$ vs. $T$ behavior in the dark state is well reproduced by a many-body microscopic model Hamiltonian which incorporates the above physical picture [9]. This is equivalent to the situation in which five singlets (S=0), nine triplets (S=1), five quintets (S=2) and one heptet (S=3), which can be formed from the six spin-½ objects, are all degenerate. The lowest energy excitation in our model corresponds to promoting one of the electrons in the Mo(IV) ion from the lower energy orbital to the higher energy orbital (Fig. 3b). The excitation energy for this configuration, when the two electrons on Mo are aligned parallel is $(U^{dd}_{Mo} + \Delta - J^{d}_{Mo} - U^{d}_{Mo})$, where $U^{d}_{Mo}$ is the Hubbard parameter for Mo 4d orbital, $U^{dd}_{Mo}$ is the inter 4d-orbital electron repulsion integral and $J^{d}$ is the direct exchange integral for the 4d orbitals. When the two electrons on Mo in this occupancy scheme are aligned antiparallel, the energy of the configuration is $(U^{dd}_{Mo} + \Delta - U^{d}_{Mo})$, relative to the ground state configuration. Assuming $J^{d}_{Mo}$ smaller than $\Delta$, the spin states arising from two different spin orientations on Mo(IV) would differ in energy by less than $\Delta$. In this case, we have two unpaired electrons on the Mo(IV) site and six unpaired electrons, one each on the six Cu(II) sites. The total number of possible spin orientations is $2^8$, and correspond to 14 singlets, 28 triplets, 20 quintets, 7 heptets and 1 nonet (S=4). The exchange pathways between Cu(II) ions and the central Mo(IV) ion are now open, leading to exchange splitting of these levels. These $2^8$ spin configurations lie within a band of energy $\sim(J^{d}_{Mo} + J_{ex}) < \Delta$, where $J_{ex}$ is the effective exchange interaction between Mo(V) and Cu(II) ions and is of the order of a few meV. However, these excited states are dipole forbidden (in a complete model this could be weakly dipole allowed) with respect to the ground state since the site charge densities in all these states are identical to that in the ground state. The next higher energy excitation corresponds to promoting both the electrons in the lower energy orbital of Mo(IV) to higher energy orbital (Fig. 3c). In this electron configuration, just as with the ground state, there is no exchange coupling between the spins and all the $2^6$ spin configurations are degenerate. Since the charge density distribution in the ground and excited states are identical, these excitations are also dipole forbidden with respect to the ground state.

The lowest energy electron configuration in which the charge density distribution on the transition metal ions is different from that in the ground state corresponds to charge transfer from the Mo(IV) ion in the ground state to any of the Cu(II) ions (Fig. 3d). This transition is strongly dipole allowed and also opens exchange pathways between the remaining Cu(II) ions and the Mo(V) ion. The exchange constant is expected to be ferromagnetic with a magnitude of a few tens of cm$^{-1}$, for the bridging cyanide ligand in 180° geometry [8]. There are a total of six different charge distributions possible for this charge transfer process and in each of these, we have five S=0 states, nine S=1 states, five S=2 states and one S=3 state. Compared to the charge-transfer excitation gap of about 3eV, the exchange splitting of these spin states is very small, of the order of a few meV. We could similarly arrive at the higher energy spin states from other possible electronic configurations. An earlier exact diagonalization study of this model gave the spectrum discussed above [9]. The transition dipole moment for electric dipole excitation from the ground state manifold to this charge transfer manifold in each of the spin spaces are comparable. Thus, all the spin states are excited with similar probability by the 3eV laser radiation. Therefore, the initial population of the excited state would not be expected to exhibit significant photomagnetism. However, the optically excited states undergo internal conversion as well as intersystem crossings due to vibronic couplings and spin-orbit interactions respectively. It is then possible that some spin states relax to the ground state while others are trapped in long-lived metastable states. The spin of all the Cu(II) ions in molecules that relax to the ground state could get decoupled and the Cu(II) centres behave as isolated spin-½ centers. Experimentally, it has been observed that the dependence of $\chi_M T$ on temperature, after irradiation, depends upon the time period of irradiation. Then, it is reasonable to expect that the S=3 excited state that is dipole connected by light can undergo internal conversion into two different types of S=3 states: one that has a slow internal conversion to the low-lying S=3 states and another which by virtue of large geometry change is not connected to the low-lying S=3 states. The formation rate of this metastable state shown in Fig. 4 could be low and once this state is formed, the decay channel for this state is only through a thermally activated process.

## 4. Kinetic Model of Photomagnetism

The above picture allows constructing a kinetic model for photomagnetism. The model involves initial laser irradiation to give a steady state population of the different spin states. After a steady state is attained, the laser irradiation is turned off and the decay of the system to the ground state is followed as a function of temperature. The kinetic equations for creating the initial steady state is given by

$$\frac{dC_p}{dt} = -4k_p C_p + k_0 C_0 + k_1 C_1 + k_2 C_2 + k_3 C_3^1$$

$$\frac{dC_0}{dt} = k_p C_p - k_{-0} C_0 - k_0^{isc} C_0 + k_{-1}^{isc} C_1$$

$$\frac{dC_1}{dt} = k_p C_p - k_{-1} C_1 + k_0^{isc} C_0 + k_{-2}^{isc} C_2 - k_{-1}^{isc} C_1 - k_1^{isc} C_1$$

$$\frac{dC_2}{dt} = k_p C_p - k_{-2} C_2 + k_1^{isc} C_1 + k_{-3}^{1,isc} C_3^1 - k_2^{isc} C_2 - k_{-2}^{isc} C_2$$

$$\frac{dC_3^1}{dt} = k_p C_p - k_{-3}^1 C_3^1 + k_2^{isc} C_2 + k_3^2 C_3^2 - k_{-3}^{1,isc} C_3^1 - k_3^1 C_3^1$$

$$\frac{dC_3^2}{dt} = k_3^1 C_3^1 - k_3^2 C_3^2 \qquad\qquad \ldots\ldots\ldots\ldots(1)$$

In the above set of equations, $C_l$ refers to the concentrations of the various species, $l=p$ corresponds to the paramagnetic ground state, $l = 0, 1, 2,$ and $3$ refer to concentrations of the excited metastable states realized upon irradiation, with spins 0, 1, 2 and 3 respectively. Although, a state with S = 4 is the highest possible spin state, we do not consider it in our model as there seems to be no experimental evidence for such a state [11]. There exist two types of S=3 states indicated by superscripts 1 and 2, the S=3 state with label 2 is de-excited only through the S=3 state with label 1 and is a thermally activated process. The rate constants $k_p$ give the rates of formation of the metastable states upon irradiation and it is taken to be the same in each spin subspace since the transition dipole moments computed for the transition from an earlier microscopic model are similar in magnitude in all the spin subspaces; $k_{-l}$ is the rate of formation of the paramagnetic state from the state with spin S = $l$; $k^{isc}_{\pm l}$ corresponds to intersystem crossing rates from a spin state with spin S = $l$ to a spin state with spin S = $l\pm1$; assuming spin-orbit interaction mechanism for intersystem crossing, the spin of a given state can change by one in a first order approximation. Although it might appear that there are a large number of model parameters, the approximate relative values of many of these can be guessed on experimental grounds. Besides, our calculations show that many of the parameters are marginal parameters which do not affect the fit qualitatively, over a wide range of their values. During the irradiation process, we assume $k_p$ = 1 and other rate constants $k_{-l}$ to be small, with $k_{-3}$ being the least. All the intersystem crossing rates $k^{isc}_{\pm l}$, are expected to be much smaller than $k_{-l}$ since spin-orbit effects would be weak in these complexes both due to partial quenching of angular momentum due to lowered symmetry, and due to smaller atomic number of the magnetic ions involved.

## 5. Results and Discussion

The Curie constant $\chi_M T$ (taken to be the magnetization) of the system is given by, $\sum C_l \chi_l T$, where $\chi_l T$ for spins 0, 1, 2, 3 and the paramagnetic species are respectively 0, 1.000, 3.001, 6.002 and 2.250 cm$^3$ K mol$^{-1}$, assuming the g value to be 2.0 [10]. The temperature dependence of the rate constants $k_l$ are obtained by using an Arrhenius dependence of the rate constant on temperature,

$$k_{-l}(T) = k_{-l}(300)\, e^{-\frac{E^l_a}{k_B}\left(\frac{1}{T} - \frac{1}{300}\right)}$$

$$k^{isc}_l(T) = k^{isc}_l(300)\, e^{-\frac{E^{l,isc}_a}{k_B}\left(\frac{1}{T} - \frac{1}{300}\right)}$$

$$k^{isc}_{-l}(T) = k^{isc}_{-l}(300)\, e^{-\frac{E^{l,isc}_a}{k_B}\left(\frac{1}{T} - \frac{1}{300}\right)} \qquad\qquad\ldots\ldots\ldots\ldots (2)$$

The rate constant $k_p$ is set to unity during irradiation process and $k_p$ is set to zero when the radiation is turned off during the relaxation process. The rate constants at 300 K and the activation energies of the model are varied to fit the experimental $\chi_M T$ vs. $T$ plots. The activation energies for the decay of spin states with S = 0, 1 and 2 ($l$ = 0, 1 and 2) are assumed to be small compared to that of the S=3 state ($l$ =

3). This appears reasonable since it has been demonstrated experimentally that photomagnetism is not observed for rigid ligands in other Mo-Cu complex analogs, showing that there is a substantial geometry relaxation around Cu(I) associated with the S=3 state [11]. The $MoCu_6$ complex has only six bridging cyano groups for the Mo ion, unlike the usual eight bridging cyano groups associated with molybdenum in the photomagnetic CuMo networks [5]. The smaller number of coordinated metallic centers and the flexibility of the blocking ligand permit geometry relaxations in the excited states. Indeed, it is also seen to be the case in Prussian Blue wherein photomagnetism vanishes when the system is too rigid with no Fe vacancy. The large geometry relaxation implies negligible internal conversion and the decay of this state is only through a thermally activated process as shown in Fig. 4. This is further reinforced by the fact that for high spin states, the energy surfaces between the ground state and the lowest excited state would be few and far leading to phonon blockage and long life-times. We could assume the activation energies for intersystem crossings to follow a similar trend though the corresponding rate constants are smaller than that for internal conversion to the ground state by approximately an order of magnitude.

To initially prepare the system, we choose $k_p = 1$, $C_p = 1$ with all other concentrations set to zero, all the rate constants evaluated at the irradiation temperature (10 K) from the room temperature values and chosen activation energies. These quantities completely define the right hand side of the kinetic equations at time t=0. The system of equations is then solved numerically to obtain the concentration of various species as a function of time until steady state concentrations are obtained. After determining the steady state concentrations, we start warming the system, in the absence of the radiation field. This is achieved by setting $k_p = 0$ in all the rate equations. The rate of warming is determined by the number of time steps spent at each temperature. At each temperature, the new rate constants are determined from the corresponding Arrhenius equations. The concentrations are evolved for the desired number of time steps by numerically solving the rate equations and the $\chi_M T$ values are computed before raising the temperature.

The best fit plots for two different sets of experimental data are shown in Fig. 5. We find that the same set of kinetic parameters fit the two experimental plots, with different irradiation times. We note that the best fit parameters correspond to large activation energy for the $S = 3$ state and they are in the order $E_a(S=3) > E_a(S=2) > E_a(S=1) > E_a(S=0)$; the rate constants at 300K follow the reverse order. The intersystem crossing rates and activation energies for intersystem crossing required to fit the data are small showing that the photomagnetism phenomena is mainly governed by the higher activation energy and low rate of internal conversion of the $S = 3$ state. When we have relatively long irradiation times, the peak height in the $\chi_M T$ vs. $T$ plot is higher. The theoretical fits for two different irradiation times are shown in Fig. 5; the upper curve corresponds to a longer irradiation time which is approximatelyseven times that of the lower curve.

To conclude, the photomagnetism in $MoCu_6$ can be modeled quantitatively using a kinetic model for relaxation of the photoexcited states with two different types of S=3 excited states, one which is connected to the low-lying states via internal conversion and another which is isolated in configuration space and connected to an S=3 excited state via a thermally activated barrier.

**Acknowledgements:** This work was carried out partly under the joint Indo-French laboratory for Solid State Chemistry supported by DST –India and CNRS – France. V.M. and C.M. are also indebted to the French Ministry of Research for funding (ACI project no. JC4123). Support from the Indo-French Centre for Promotion of Advanced Research (IFCPAR) under the project 3108-3 is also gratefully acknowledged. SR and RR thank Professor Michel Verdaguer for inspiring discussions.

**Figure captions:**

Fig. 1: (Color Online) Structure of *MoCu₆* complex. Yellow sphere corresponds to *Mo* ion and blue spheres correspond to *Cu* ions.

Fig. 2: $\chi_M T$ vs. *T* behaviour of MoCu₆ complex before irradiation (filled circles) and after 1 hour (open squares) and after 7 hour (open circles) of irradiation.

Fig. 3: (Color Online) Occupation of orbitals in ground state and different excited state manifolds. (a) The ground state manifold (b) The lowest energy excited state manifold (c) The second lowest energy excited manifold and (d) the lowest energy charge-transfer excitation manifold.

Fig. 4: (Color Online) Schematic representation of the kinetic processes of Eqn. (1) in the text.

Fig. 5: Open squares and open circles are the experimental plots of $\chi_M T$ vs. *T* plots for irradiation times of 7 hours and 1 hour respectively. Solid lines correspond to the calculated $\chi_M T$ values from our model. Best fits are obtained for the following parameters: (*i*) Activation energies for internal conversion: $E_a^0 = 18.5$; $E_a^1 = 20.5$; $E_a^2 = 39.0$; $E_a^3 = 100.0$; $E_3^1 = 16.7$; $E_3^2 = 22.7$; (*ii*) Rate constants for internal conversion at 300 K: $k_{-0} = 0.6; k_{-1} = 0.48; k_{-2} = 0.11; k_{-3} = 0.0002$; $k_3^1 = 0.0026$; $k_3^2 = 0.0013$; (*iii*) Activation energies for intersystem crossing: $E_a^{0,isc} = 10.4$; $E_a^{1,isc} = 10.5$; $E_a^{2,isc} = 10.7$; $E_a^{3,isc} = 16.7$; (*iv*) Rate constants for intersystem crossing at 300 K: $k_0^{isc} = 0.43$; $k_1^{isc} = 0.33$; $k_2^{isc} = 0.21$; $k_3^{isc} = 0.0026$; $k_{-0}^{isc} = 0.43$; $k_{-1}^{isc} = 0.33$; $k_{-2}^{isc} = 0.21$; $k_{-3}^{isc} = 0.0026$. All the energies are in units of $k_B T$ and all the rate constants are in arbitrary units of inverse time. The calculated upper curve corresponds to a longer period of irradiation (7.4 times longer than that of the lower curve).

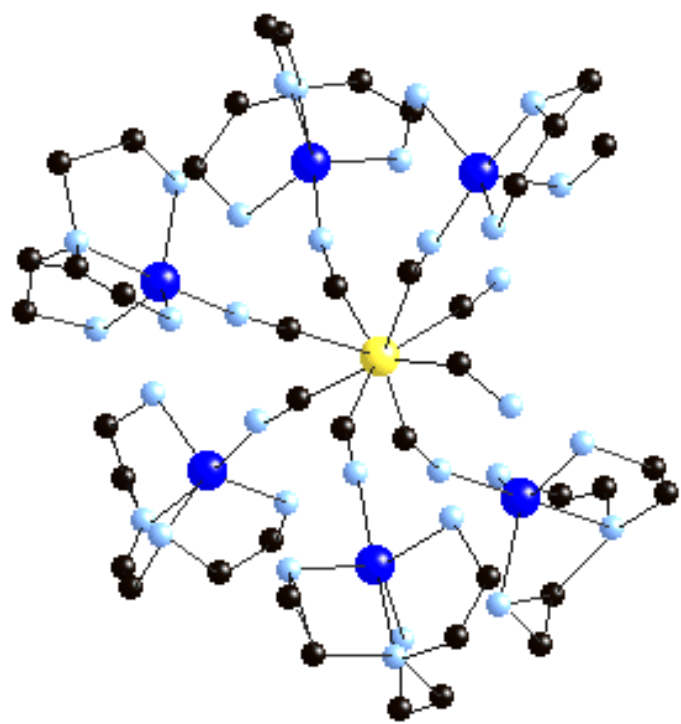

Fig. 1
13

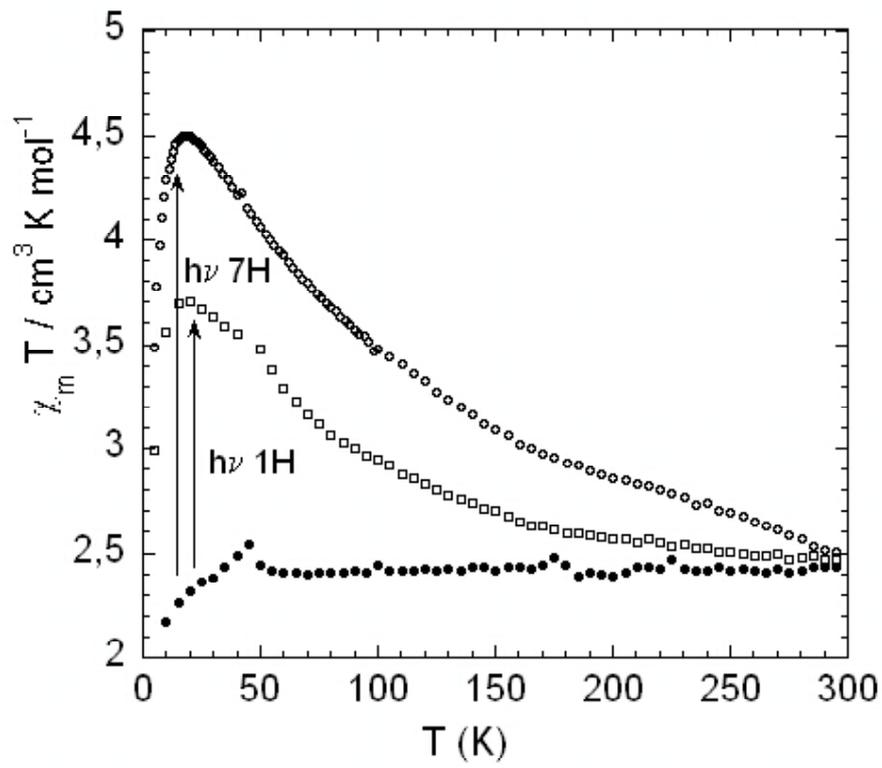

Fig.2

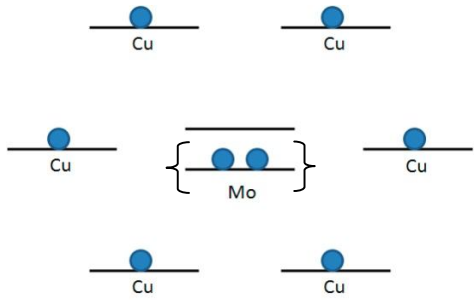
(a)

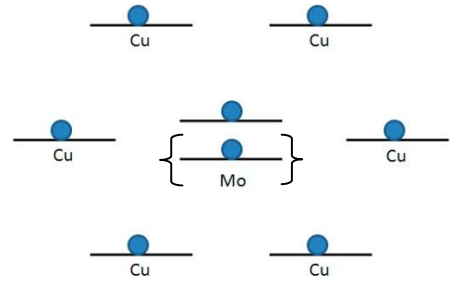
(b)

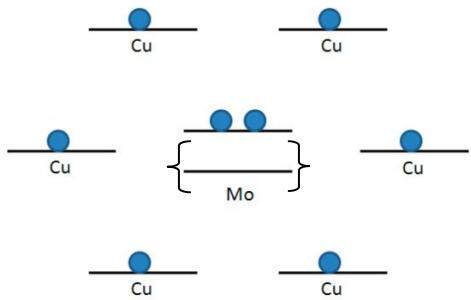
(c)

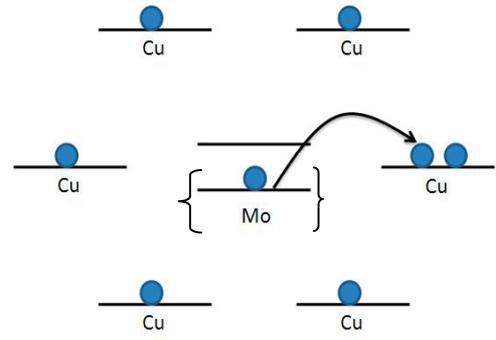
(d)

Fig. 3

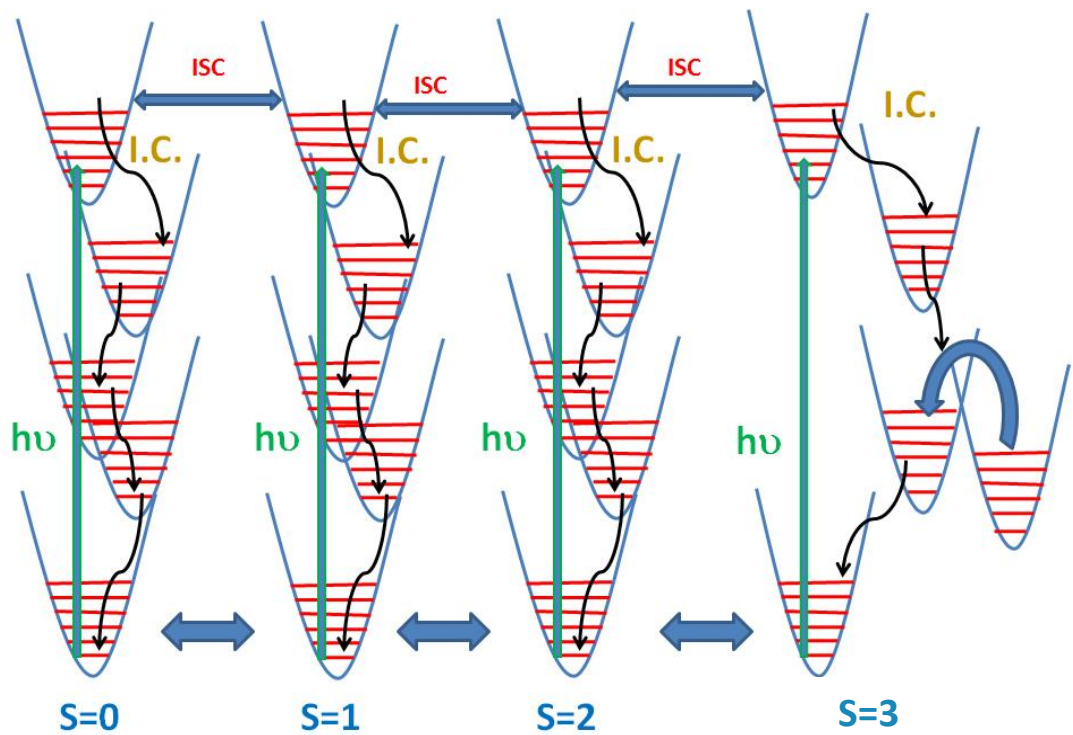

Fig. 4

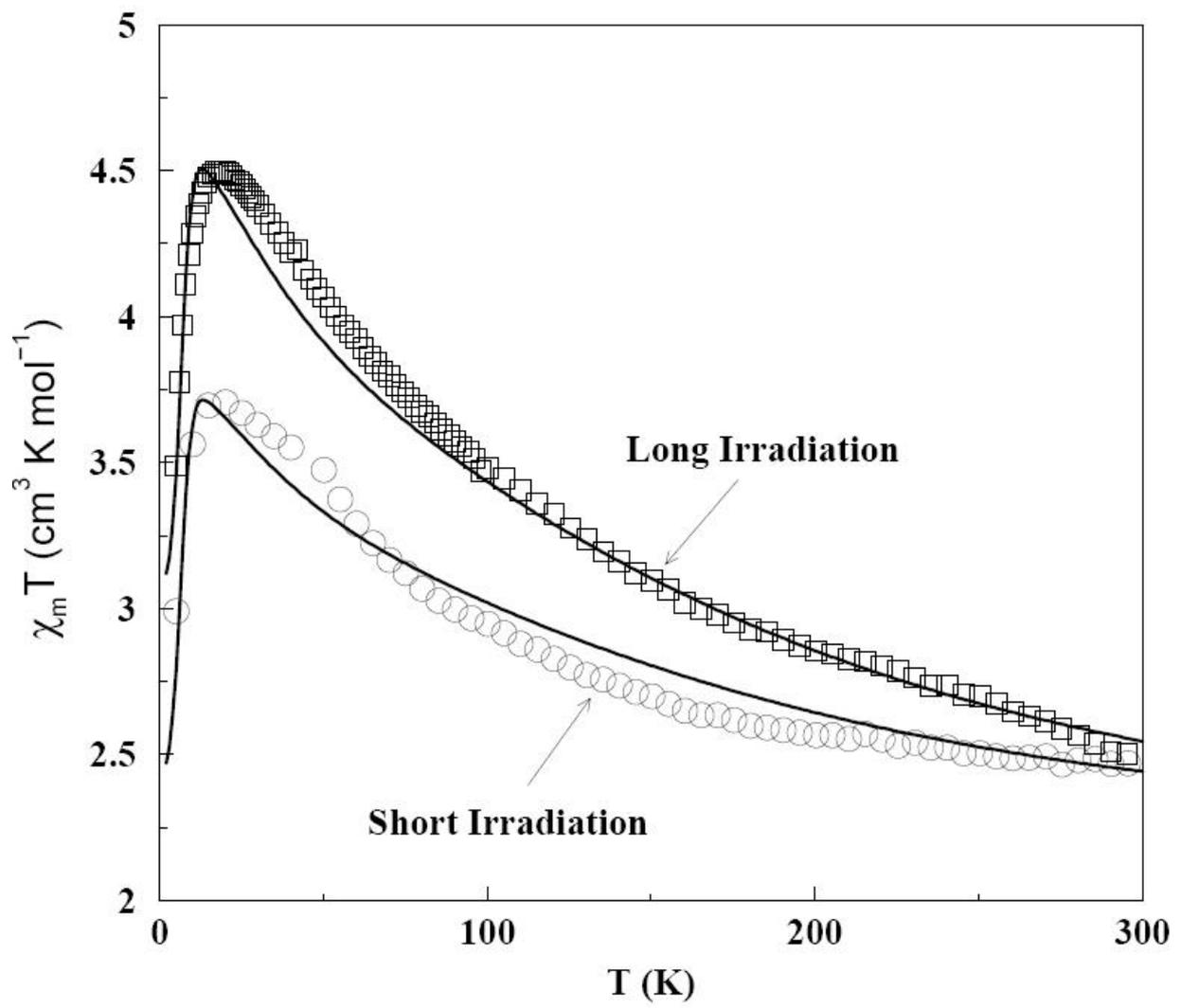

Fig. 5